\documentclass[12pt]{article}

\newcommand{\be}{\begin{equation}}
\newcommand{\ee}{\end{equation}}
\newcommand{\ba}{\begin{eqnarray}}
\newcommand{\ea}{\end{eqnarray}}

\begin{document}
\title{\bf A note on Gaussian integrals over 
paragrassmann variables}
\author{
Leticia~F.~Cugliandolo$^{a,b}$, G.~S.~Lozano\thanks{Conicet}$~^c$,
\\
E.~F.~Moreno$^{*d}$ and
F.~A.~Schaposnik\thanks{Associated with CICPBA}$~^d$
\vspace{0.2 cm}
\\
{\normalsize \it $^a$Laboratoire de Physique Th\'eorique de l'Ecole 
Normale
Sup\'erieure}
\\
{\normalsize \it 24 rue Lhomond, 75231 Paris C\'edex 05, France}
\\
{\normalsize \it $^b$Laboratoire de Physique Th\'eorique et Hautes 
Energies
Jussieu}
\\
{\normalsize\it Tour 16, 1er \'etage,
4 Place Jussieu, 75252 Paris C\'edex 05, France
}
\\
{\normalsize\it $^c$Departamento de F\'\i sica, FCEyN, Universidad de
Buenos Aires}\\
{\normalsize\it Pab.1, Ciudad Universitaria,
Buenos Aires, Argentina.}\\
{\normalsize\it  $^d\,$Departamento de F\'\i sica, Universidad 
Nacional
de La Plata}\\
{\normalsize\it C.C. 67, 1900 La Plata, Argentina.}
}

\date{\hfill}
\maketitle

\begin{abstract}
We discuss the generalization  of the 
connection between the determinant of an operator entering 
a quadratic form and the associated Gaussian
path-integral valid for grassmann variables
to the paragrassmann case
[$\theta^{p+1}=0$ with $p=1$ ($p>1$) for grassmann (paragrassamann)
variables].
We  show that the $q$-deformed commutation
relations of the 
paragrassmann variables lead naturally to consider
$q$-deformed
quadratic forms related to multiparametric
deformations of $GL(n)$
and their corresponding $q$-determinants. We suggest a
possible
application to the study of disordered
systems.
\end{abstract}
\date{\hfill}

\newpage

Using anticommuting functions as integration variables, Matthews
and Salam showed that the path-integral
for  a system of relativistic fermions in an external field gives
the determinant of the Dirac operator  \cite{MS}. That is, the
fermionic partition function does not yield a negative power of the
determinant, as in the bosonic case, but a positive power, $p= +1$. 
Ten
years later, Berezin completed his analysis of noncommutative algebras
and fermion systems, making clear that the natural framework to
define fermionic path-integrals was that of Grassmann algebras
 \cite{B}.

The main ingredient behind the result
\begin{equation}
\int \prod_{i=1}^n d\theta_i d\bar\theta_i
\exp\left( \bar\theta_i A_{ij} \theta_j \right)
= \det A
\label{grass}
\end{equation}
when one integrates over two sets of $n$
Grassmann variables $\theta_i$ and
$\bar\theta_i$ are the anticommutation rules
\be
[\theta_i,\theta_j]_+=0 \; ,
\;\;\;\;\;\;\;
[\bar\theta_i,\bar\theta_j]_+=0
\; ,
\ee
for all pairs $i$,$j$, that imply $\theta_i^2 = \bar\theta_i^2 =0$
for all $i$. In Eq.~(\ref{grass}) we neglected an irrelevant
factor related to the definition of the integration measure. The
summation rule over repeated indices is used henceforth. Notice
that the condition
$ [\bar\theta_i,\theta_j]_+=0, $
which is usually imposed, is not necessary for the validity of
Eq.~(\ref{grass}). In fact, a relation of the form
$ \bar\theta_i \theta_j= \alpha\; \theta_j
\bar\theta_i 
\; ,
$
with $\alpha$ some $c$-number, yields the same result 
(modulo an irrelevant
normalization).

In order to construct a path-integral representation of the $p$-th
power of a determinant it seems natural to use $p$-paragrassmann
variables such that
\be
\theta_i^{p+1} = \bar\theta_i^{p+1} = 0
\; ,
\;\;\;\; \mbox{for all} \; i \; .
\label{cero}
\ee
Consistent integration rules  for $\theta_i$
and $\bar\theta_i$ take the form~(see for instance 
\cite{Baulieu,Matheus-Valle:qs,Durand})
\be
\int d\theta_i \; \theta_i^r = {\cal N} \; \delta_{r,p} \; , 
\;\;\;\;\;\;
\int d\bar \theta_i \; \bar{\theta}_i^r = { \bar{\cal N}} \; 
\delta_{r,p}
 \; ,
\label{key}
\ee
where ${\cal N}$ and $\bar{\cal N}$ are two complex numbers
that, without loss of generality, we set to $1$.
Indeed, one can easily see that the Gaussian
integral of a {\it diagonal} form,
$\bar\theta_i A_{ij} \theta_j = \bar\theta_i \lambda_i
\theta_i$,
 that is quadratic in the
pair of $p$-paragrassmann variables $\theta_i$ and $\bar\theta_i$,
leads to the $p$-th power of
the product of the diagonal elements,
\begin{eqnarray}
\int \prod_{i=1}^n d\theta_i d\bar\theta_i 
\exp\left( \bar\theta_i A_{ij} \theta_j \right)
= \left(\prod_{i=1}^n \lambda_i \right)^p = (\det A)^p
\; .
\label{pgrass}
\end{eqnarray}
%
Here and in what follows we use the ordinary definition of the 
exponential,
$e^x \equiv \sum_{m=0}^\infty x^m/m!$.

However, contrary to what seems to be accepted in the literature,  
it is not straightforward to obtain an analogous result
whenever the quadratic form is not diagonal. The reason is that
the change of variables needed to bring $A$ to a diagonal form,
spoils the commutation rules of the paragrassmann variables,
unless $p=1$. Thus, in order to define a consistent path-integral
for paragrassmann variables,  one has to take into account
paragrassmann changes of variables (a fact that, to our knowledge,
has not been discussed in the literature
\cite{Matheus-Valle:qs}-\cite{Rausth}).

It is the purpose of this work to fill this gap by developing a
consistent framework to integrate paragrassmann variables
that allows one to deal with Gaussian integrals of non-diagonal
quadratic forms.  We shall see that the quantum
group $GL_{q,q'}(n)$ (with $qq'$ a primitive root of unity)
enters naturally into play.

Let us start by fixing the commutation rules for the $\theta_i$
variables among themselves. For simplicity we consider just two 
variables
$\theta_1$ and $\theta_2$ and impose the following $q$-commutation
rule
\be
\theta_1 \cdot \theta_2 = q\; \theta_2 \cdot \theta_1
\; ,
\label{5}
\ee
with $q$ a  $c$-number. 
Usually, see
for instance \cite{Isaev}, this number is taken to be a
primitive root of unity,
%
$q^{p+1}=1$, $q^m \neq 1$ for all $m < p+1$,
%
but this condition is not necessary to define a consistent integral
and we shall not impose it.

We now consider a linear change to new variables
\ba
\omega_1 &=& a\; \theta_1 + b\; \theta_2
\; ,
\nonumber
\\
\omega_2 &=& c\; \theta_1 + d\; \theta_2
\; ,
\label{cambio}
\ea
with $a, b, c, d$ certain in principle non-commuting
parameters that  commute with $\theta_1$ and $\theta_2$ and that 
we encode in a $2\times 2$ matrix 
\be   A = \left(
\begin{array}{ll}
a & b\\
c & d
\end{array}
\right) \; .
\label{matrixA}
\ee
Suppose that
\ba
&& a\cdot b = q'\; b \cdot a 
\; ,
\nonumber \\
&& c\cdot d = q'\; d \cdot c 
\; ,
\nonumber
\ea
with $q'$ a second complex parameter.
We want the $\omega$'s to have the same commutation properties as
the $\theta$'s.
It is easy to show that, defining
\be 
z=qq' 
\; , 
\ee
we obtain
\be (a\, \theta_1 + b \, \theta_2)^m = \sum_{l=0}^m {m\choose
l}_{\!\!z}   b^{m-l} \cdot a^{l} \; \theta_2^{m-l} \cdot
\theta_1^{l} \label{binomio} \ee
with
\[
{m\choose l}_{\!\!z}
\equiv \frac{(m)_{z}!}{(l)_{z}! (m-l)_{z}!}
\; ,
\;\;\;\;\;\;\;\;\;\;\; (m)_{z} \equiv
\frac{1-z^m}{1-z} \; ,
\]
\[
(m)_z! = (m)_z (m-1)_z! \; , \;\;\;\;\;\;\;\;\;\;\;\;\;\;\;\; 1_z=1 
\; .\;\;\;\;\;\;\;\;\;\;\;
\]
Then, conditions
\be
\omega_1^{p+1}= \omega_2^{p+1}=0
\label{omegap}
\ee
require  $z$ to be a primitive root of unity,
\be
z^{p+1}=1 \; , \;\;\;\;\;\;\; z^m \neq 1 \;\;\; \mbox{for all}\;\; m < 
p+1 \; .
\label{z}
\ee
Note that $q$ and $q'$ are not fixed separately to be roots of unity.

If we further enforce the analogous to condition
(\ref{5}),
\[
\omega_1 \cdot \omega_2 = q\; \omega_2 \cdot \omega_1
\; ,
\]
the following commutation rules between the coefficients in the
change of variables (\ref{cambio}) should hold
\ba
i) && a\cdot c = q\; c \cdot a
\; ,
\nonumber\\
ii) && b\cdot d = q\; d \cdot b
\label{8}
\; ,\\
iii)&& a\cdot d - d\cdot a  - q'\; c \cdot b + 1/q\; b \cdot c=0
\nonumber
\; .
\ea
Already at this point we see that $a,b,c,d$ cannot be ordinary
$c$-numbers. Indeed, this would imply $q=q'=1$ but condition
(\ref{z}) requires $(qq')^m \ne 1$ for any $m<p+1$.

Conditions (\ref{8}) do not exhaust the commutation rules
among the coefficients $a, b, c, d$. If we impose 
\be
b\cdot c = q/q'\; c\cdot b
\; ,
\ee
relation $iii)$ becomes
\be
iii) \;\;\;\;\;\; a\cdot d - d\cdot a  = (q'-1/q)\; b \cdot c=
(q-1/q')\; c \cdot b \; .
\ee
The resulting commutation rules imply that
 $A$ belongs to the quantum group $GL_{q,q'}(2)$ 
(some of its properties are discussed in Appendix A).

At this point we can compute the Jacobian  $J$ associated with
the transformations (\ref{cambio}) via
\be d\omega_1 \cdot d\omega_2 = J d\theta_1 \cdot d  \theta_2
\; .
\label{jota}
\ee
 Since 
$\omega_1$ and $\omega_2$ are $p$ paragrassmann variables 
their integration yields 
\be
\int d\omega_1 \cdot d\omega_2 \; \cdot \omega_1^{r} \cdot \omega_2^{s} 
=
\delta_{r p}\delta_{sp}
\label{normal}
\ee
where we have omitted the overall normalization constant. Consider
now the integral
\begin{eqnarray}
1=
\int d\omega_1 \cdot d\omega_2 \cdot \omega_1^{p} \cdot \omega_2^{p} =
J\,\int d\theta_1 \cdot d\theta_2 \cdot (a\; \theta_1 + b \;
\theta_2)^{p} \cdot (c\; \theta_1 + d\; \theta_2)^{p}
\; .
\label{sinn}
\end{eqnarray}
The inverse of the Jacobian is then given by
\be
J^{-1} = \int d\theta_1 \cdot d\theta_2 \cdot (a\; \theta_1 + b\;
\theta_2)^{p} \cdot (c\; \theta_1 + d\; \theta_2)^{p} \equiv {\cal I}
\; .
\label{sinnn}
\ee
Using the expression (\ref{binomio}) and the integration rules
(\ref{key}) it is easy to show that the only contribution to the
integral ${\cal I}$ 
comes from terms having $p$ powers of $\theta_1$ and $p$ powers of
$\theta_2$,
\begin{eqnarray}
{\cal I}&=& \sum_{l=0}^{p} {p \choose l}_{\!\!z}^2 b^{p-l} \cdot a^l
\cdot d^{l} \cdot c^{p-l}
\int d\theta_2 \cdot d\theta_1 \cdot
\theta_2^{p-l} \cdot \theta_1^{l} \cdot \theta_2^{l} \cdot 
\theta_1^{p-l}
\nonumber\\
&=& \sum_{l=0}^{p}  \Gamma^p_l \; a^l \cdot d^l \cdot b^{p-l} \cdot 
c^{p-l}
\label{p1}
\end{eqnarray}
with
\be
\Gamma^p_l \equiv \frac{1}{z^{l(p-l)}}{ p \choose l}^2_{\!\!z}
q^{-(p-l)^2} \label{zz}
\; .
\ee
Here we have used the commutation rules given above for $a,b,c,d$
and normalized the integral of the $\theta$ and $\bar \theta$
variables as in Eq.~(\ref{normal}). After some algebra, 
Eq.~(\ref{p1}) can be accommodated  as
\be
{\cal I} = (a \cdot d - q'\; b \cdot c)^p \equiv \Delta^p
\; .
\label{ys}
\ee
Indeed, writing
\be
(a \cdot  d - q' \; b \cdot c)^p
= \sum_{l=0}^p \Lambda_l^p  \; a^l \cdot d^l \cdot b^{p-l} \cdot  
c^{p-l}
\label{b1}
\ee
we can show that the coefficients $\Lambda_l^p$ coincide with the
coefficients $\Gamma_l^p$ in Eqs.~(\ref{p1}) and (\ref{zz}). To this
end, using $\Delta\equiv a \cdot  d - q'\; b \cdot c$ and
Eq.~(\ref{conmudet}) we have 
\be
\Delta^{p+1} = \Delta^p \cdot ( a \cdot d - q' \; b \cdot c)
= a \cdot \Delta^p \cdot d - q' \Delta^p \cdot b \cdot c
\; .
\label{b2}
\ee
and replacing $\Delta^p$ with the expression in Eq.~(\ref{b1})
we obtain
%
\be
\Delta^{p+1} = \sum_{l=0}^{p+1} a^l \cdot d^l \cdot b^{p +1-l} \cdot 
c^{p+1-l}
\; z^{p+1-l} \;
\left( \Lambda_{l-1}^p - q^{2l - 2p - 1}
\Lambda_{l}^p
\right)
\label{b3}
\ee
where we defined $\Lambda_{-1}^p = \Lambda_{p+1}^p = 0$.
Equation~(\ref{b3}) determines the following recurrence relation for 
$\Lambda$:
\be
\Lambda^{p+1}_l  = z^{p+1-l}
\left( \Lambda_{l-1}^p - q^{2l - 2p - 1}
\Lambda_{l}^p
\right)
\; ,
\label{b4}
\ee
with the initial conditions
%
$
\Lambda_0^1 = - q'$, $\Lambda_1^1 = 1
$ 
%
that is solved by
\be
\Lambda^p_l = (-1)^{p-l} q^{-(p-l)^2} z^{(p-l)(p-l+1)/2}
{p \choose l}_{\!\!z}
\; .
\label{b6}
\ee
When $z$ is a primitive root of unity we have
\be
{p \choose l}_{\!\!z} = (-1)^l z^{-l(l+1)/2}
\; ,
\label{b7}
\ee
and Eq.~(\ref{b6}) becomes
\be \Lambda^p_l = \frac{1}{z^{l(p-l)}}{ p \choose l}^2_{\!\!z}
q^{-(p-l)^2} =\Gamma^p_l \label{b8} \ee
completing the proof of the identity in Eq.~(\ref{ys}).

Thus, the Jacobian of 
the linear change of variables (\ref{cambio}) that is associated to an
element $A \in GL_{q,q'}(2)$,
is given by the inverse $p$-th power of the
``$q$-determinant'' of $A$:
\be J=(\det A)^{-p}
\label{inter}
\ee
with the generalized determinant defined as
\be
\det A \equiv  a \cdot d-q'\; b \cdot c
\label{deter}
\; .
\ee
%
%
%

We shall now discuss the calculation of the Gaussian integral for a non 
diagonal quadratic form.
In order to do it, we need to  fix the complete paragrassmann 
algebra by defining  the commutation rules between the
$\bar \theta$'s and the $\theta$'s and those among different $\bar
\theta$'s.

The simplest possibility is  to demand that {\sl independent}
$GL_{q,q'}(2)$ transformations for the $\theta$'s and $\bar
\theta$'s preserve the commutation relations, leading to the $\bar
\theta$'s and $\theta$'s commuting   with each other
independently of their indices (up to a factor which for
simplicity we take to be 1). That is
\be
\bar \theta_i \cdot \theta_j = \theta_j \cdot \bar \theta_i
\;, \;\;\;\; \forall \,i,j
\label{nombre}
\ee
(notice that these conditions are not those imposed in
Ref.~\cite{Isaev}).

Regarding the commutation relations for the ${\bar \theta_i}$'s, 
we have two possible choices,
\begin{eqnarray}
C1 :&&\;\;\; {\bar \theta}_1 \cdot {\bar \theta}_2 = q^{-1}\; {\bar 
\theta}_2
\cdot {\bar \theta}_1 
\; ,
\label{c1}\\
C2 :&&\;\;\; {\bar \theta}_1 \cdot {\bar \theta}_2 = q\; {\bar 
\theta}_2
\cdot {\bar \theta}_1 
\; .
\label{c2}
\end{eqnarray}
Let us  start by analyzing $C1$. Consider the integral
\be
I = \int d \bar \theta_1 \cdot d\bar \theta_2 \cdot d \theta_1 \cdot 
d\theta_2
\cdot
e^{{\bar \theta}_i A_{ij}
\theta_j}
\label{integral}
\ee
where $A$ is a matrix belonging to $GL_{q,q'}(2)$. Changing the
integration variables as
\begin{eqnarray}
\theta_i \to \omega_i =A_{ij} \theta_j
\; ,
\;\;\;\;\;\;\;\;
%
\bar \theta_i \to \bar \theta_i
\; ,
\end{eqnarray}
the result (\ref{inter}) for the Jacobian and the fact that 
$\omega_i$ and $\bar\theta_j$ commute lead to  
\be I = (\det A)^p\;\int
d\bar \theta_1 \cdot  d\bar \theta_2 \cdot d \omega_1 \cdot  d\omega_2
\cdot
e^{{\bar \theta_i} \cdot \omega_i}
\label{for1}
%
=
(\det A)^p
\ee
%
whenever $qq'  = 1$ is a primitive root of
unity.

Since for the rules $C1$ the parameters  $q$ and $q^{-1}$ play a
dual role regarding the $\theta$'s and $\bar \theta$'s, it is
natural to consider, apart from the case discussed above, the one
corresponding to quadratic forms $A \in
GL_{q'^{-1},\,{q^{-1}}}(2)$ (notice that $q'^{-1} q^{-1}$ is also
a primitive root of unity; the reason why the order of the
deformation parameters is transposed will become clear
immediately).
The appropriate change of variables is in this case
\begin{eqnarray}
\theta_i \to  \theta_i
\; ,
\;\;\;\;\;\;\;\;
%
\bar \theta_i \to \bar \omega_i = \bar \theta_j A_{ji}
\; .
\end{eqnarray}
This is a consistent change of variables only for $A^T \in
GL_{{q^{-1}},\,q'^{-1}}(2)$ or $A \in GL_{q'^{-1},{q^{-1}}}(2)$
with $q'q$ a primitive root of unity. An argument similar to
the one leading to Eq.~(\ref{for1}) yields  also in this case
\be
\int
d\bar \theta_1 \cdot  d\bar \theta_2 \cdot d \theta_1 \cdot  d\theta_2
\cdot  e^{{\bar \theta}_i \cdot A_{ij} \cdot \theta_j} = (\det A)^p .
\label{for2}
\ee

Now, let us consider the case $C2$ in which the commutation relations
for the $\bar \theta$'s are identical to those among the $\theta$'s
The integral of the diagonal form 
\be
\int d \bar \theta_1 \cdot  d\bar \theta_2  \cdot d \theta_1 \cdot 
d\theta_2
\; e^{\bar \theta_1 \cdot \theta_1 + \bar \theta_2 \cdot \theta_2}
\label{diagonal}
\ee
vanishes.
Moreover, with an argument similar to the one used in the case $C1$ 
one proves that any integral of the form (\ref{integral}) with
$A \in GL_{q,q'}$ (with $qq'$ a primitive root of the unity) also
vanishes. We can however find a non trivial result by noticing that
$C2$ differs from $C1$ only by the exchange
\ba
{\bar \theta_1} \rightarrow {\bar \theta_2} \; , \;\;\;\;\;\;
{\bar \theta_2} \rightarrow {\bar \theta_1}
\; .
\ea
If we construct a quadratic form 
${\bar \theta}_i \cdot K_{ij} \theta_j$
with a $2\times 2$ matrix $K$ 
with entries satisfying
%
%
%
%
%
\be
\begin{array}{rcl}
a\cdot b &=& q'^{-1}\; b \cdot a \; ,\\
a\cdot c &=& q\; c \cdot a \; ,\\
b\cdot d &=& q\; d \cdot b \; ,\\
c\cdot d &=& q'^{-1}\; d \cdot c \; ,\\
a\cdot d &=& q/q'\; d \cdot a \; ,\\
b\cdot c - c\cdot b &=& (q'-1/q)\; a \cdot d \; , \\
\end{array}
\label{matrices}
\ee
we have 
\be \int d\bar \theta_1 \cdot  d\bar \theta_2 \cdot d \theta_1
\cdot  d\theta_2 \cdot e^{{\bar \theta}_i \cdot K_{ij} 
\theta_j} = (\det K)^p 
\label{for3}
\ee
where
\be
\det K\equiv a\cdot d -q\; c\cdot b
\; .
\ee
The elements of the matrices $ K$ satisfy 
commutation relations preserved under
simultaneous left $GL_{q q'}(2)$ and right $GL_{q^{-1}
q'^{-1}}(2)$ rotations:
\be \label{k-comm-1}
K_{ij} \to K'_{ij} = M_{il} {\bar M}_{sj} K_{ls}  \; ,
\;\;\;\; M\in GL_{q q'}(2),\; \bar M \in GL_{q^{-1} q'^{-1}}(2)
\ee
where $[M_{ij},\bar M_{lm}]=[M_{ij},K_{lm}]=[\bar M_{ij}, K_{lm}]=0$ 
\cite{Azcarra}.
Notice nevertheless 
that, unlike the $A$ matrices in $GL_{q,q'}(2)$, the $K$ matrices
cannot be diagonal. This is consistent with our previous statement
that commutation rules $C2$ are incompatible with
integration of diagonal quadratic forms.

So far 
we have considered an algebra with two 
paragrassmann variables. As already mentioned in 
\cite{Filippov}  the analysis of these algebras becomes rather involved 
as $n$ increases and,
 to our knowledge, a complete classification is still missing. 
Nevertheless, we believe that the generalization of our 
result to $n>2$ will lead us to consider $q$-commutation  of the form
\ba
\theta_i \cdot \theta_j &=& 
R^{(1)}_{ij,kl} \; \theta_k \cdot \theta_l \\
{\bar \theta_i} \cdot {\bar\theta_j} &=& 
R^{(2)}_{ij,kl} \; {\bar \theta_k} \cdot {\bar \theta_l}
\ea
where $R^{(1)}$ and $R^{(2)}$ are matrices 
related to multiparametric deformations of 
$GL(n)$. We expect to report on this issue in the future.


In conclusion, in this paper we showed how to introduce consistent 
Gaussian
integrals over paragrassmann variables. Surprisingly, one is
obliged to introduce  elements of quantum groups in the quadratic
forms to allow for linear changes of variables in the integration.

Even if rather abstract at face value this result is a first step
in our program to extend the supersymmetric approach
 to disordered
systems in a way that its relation with the replica method becomes
more general and transparent. In brief, many interesting problems are
represented with ``disordered''  field theories in which some
parameters are taken from a probability distribution (these can be
random exchanges, masses, fields, etc.) In general, one is
interested in knowing their averaged properties, {\it i.e.} the
behavior of observables averaged over disorder
\begin{equation}
[ O ] \equiv \left[ \left. -\frac1{Z} \frac{\partial Z}{\partial h}
\right|_{h=0} \right]
\end{equation}
with
%
$Z  \equiv \int d fields \; \exp\left(-S[fields,disorder]-h O\right)$
%
and the square brackets representing the average over the
distribution of random parameters. A typical example is given by
the calculation of the averaged spectral properties of random
matrices.
The replica and supersymmetric methods
allow one to represent the normalization $1/Z$ in exponential
form. In the former one {\it replicates} the system by making
$p-1$ identical copies of it and writes~\cite{Mepavi}
\begin{equation}
\frac{1}{Z} =\lim_{p\to 0} Z^{p-1} = \lim_{p\to 0} \; (\det 
A)^{-(p-1)/2}
\; ,
\end{equation}
with the latter identity holding for a Gaussian model. In the
latter, for a Gaussian problem, one writes~\cite{Efetov}
\begin{equation}
\frac{1}{Z} =(\det A)^{1/2} = \frac{\det A}{(\det A)^{1/2}}
=
\int \prod_{i} d\phi_i d\bar\theta_i  d\theta_i \;
e^{\bar\theta_i A_{ij} \theta_j +\phi_i A_{ij} \phi_j}
\end{equation}
with $\bar\theta_i$ and $\theta_i$ Grassmann  and $\phi_i$ real 
bosonic
variables. In both cases one takes advantage of the 
thermodynamic large 
$n$
limit to analyze the effective replicated real bosonic and
supersymmetric field theories.
The connection between the two methods has not been fully
clarified yet. However, ``mappings'' between the
replica expressions when $p\to 0$ and the supersymmetric
ones are easy to construct~\cite{jorge}.
A trivial example is $\lim_{p\to 0}
\sum_{k=1}^p 1= 0 =\int d\theta d\bar \theta$. (Indeed, one can trace 
the
relation to the properties of the $0$-dimensional replica space
and superspace.) A clue to the connection 
between the two approaches might come from the development of an 
extended supersymmetric treatment that relates to the replica one 
{\it for finite $p$}. This may also make possible the computation 
of some interesting
properties that need manipulations of the
finite $p$ replica expressions (sample-to-sample fluctuations
being one such example).
A natural way of representing 
the $(p-1)$-th power of a determinant is to introduce copies of the 
fermionic variables. Another, as we showed here, is to use 
variables with extended statistics. 
 We expect to 
report on progress in the development of an extended 
supersymmetric approach to the study of problems with quenched 
disorder
in a forthcoming publication.

\vspace{0.2cm}

\noindent{\underline{Acknowledgments}}
\newline
We thank J. Kurchan, R. Monasson and G. Semerjian for very useful 
discussions.
We acknowledge financial support from a CNRS-Conicet
collaboration, an Ecos-Sud grant and Fundaci\'on Antorchas.
L. F. C. and G. S. L. are research associates at ICTP - Trieste. L. F. .
is a Fellow of the Guggenheim Foundation. L.
F. C. thanks the Universities of Buenos Aires and La Plata and Harvard
University, and G. S. L. and F. A. S. thank the LPTHE - Jussieu for
 hospitality during the preparation of this work.

\section*{Appendix A: Quantum group $GL_{q,q'}(2)$}

Let us recall some properties of the quantum group
$GL_{q,q'}(2)$ :

\begin{itemize}
\item[1.] {\sl Closure under co-multiplication}\\
If
\[
g=\left(
\begin{array}{rr}
a & b\\
c & d
\end{array}
\right) \; , \;\;\;\;\;\;\;\;\;
g'=\left(
\begin{array}{rr}
a' & b'\\
c' & d'
\end{array}
\right)
\]
are elements of $GL_{q,q'}(2)$ such that the entries $a,b,c,d$
{\it commute} with the entries $a',b',c',d'$, then the product
\[
g \cdot g'=\left(
\begin{array}{rr}
a \cdot a' + b \cdot c'& a \cdot b' + b \cdot d'\\
c \cdot a' + d \cdot c' & c \cdot b' + d \cdot d'
\end{array}
\right)
\]
also belongs to $GL_{q,q'}(2)$

\item[2.] {\sl Existence of the inverse (antipode)}\\
The element
\[
g^{-1}=\left(
\begin{array}{rr}
d & -b q'^{-1}\\
-c q' & a
\end{array}
\right) \cdot \Delta^{-1}
\]
where
\be \Delta \equiv  a \cdot d - q'\; b  \cdot c \ee
is the inverse of $g$ and is an element of
$GL_{q^{-1},q'^{-1}}(2)$.

\item[3.] {\sl Determinant}\\
The object
%
$\Delta = a \cdot d - q'\; b  \cdot c = a \cdot d - q\; c \cdot b =
d \cdot a - q^{-1}\; b \cdot c =
d \cdot a - q'^{-1}\; c \cdot b
$
%
is defined as the {\sl determinant} of $g$. It   satisfies
\be
\Delta \cdot \left(
\begin{array}{rr}
a &  b \\
c  & d
\end{array}
\right) 
=
\left(
\begin{array}{rr}
a & q/q'b \\
c & q'/q d
\end{array}
\right) \cdot \Delta
\label{conmudet}
\ee
\item[4.] The element $g^m$ belongs to $GL_{q^m,q'^m}(2)$.
\end{itemize}


\begin{thebibliography}{99}

\bibitem{MS}   
P. T.~Matthews and A. Salam, Il Nuovo Cim. {\bf XII} (1954) 563,
{\it ibid}
{\bf 2} (1955) 120.

\bibitem {B} F. A. Berezin,
{\it The Method of Second Quantization},
(New York, Academic Press, 1966).

\bibitem{Baulieu}
 L. Baulieu and E. Floratos, 
Phys.\ Lett.\ B {\bf 258} (1991) 271.

\bibitem{Matheus-Valle:qs}
J.~L.~Matheus-Valle and M.~A.~Monteiro,
Phys.\ Lett.\ B {\bf 300} (1993) 66.

\bibitem{Durand}
S.~Durand,
Mod.\ Phys.\ Lett.\ A {\bf 8} (1993) 2323.

\bibitem{Filippov:wk}
A.~T.~Filippov, A.~P.~Isaev and A.~B.~Kurdikov,
Mod.\ Phys.\ Lett.\ A {\bf 7} (1992) 2129.


\bibitem{Durand:1993zp}
S.~Durand,
Phys.\ Lett.\ B {\bf 312} (1993) 115.

\bibitem{Durand:1993zt}
S.~Durand,
Mod.\ Phys.\ Lett.\ A {\bf 8} (1993) 1795.



\bibitem{Fleury:1995ca}
N.~Fleury and M.~Rausch de Traubenberg,
Mod.\ Phys.\ Lett.\ A {\bf 11} (1996) 899.

\bibitem{Isaev}
A.~P.~Isaev,
Int.\ J.\ Mod.\ Phys.\ A {\bf 12}  (1997) 201.

\bibitem{Rausth}
For a review and reference to the original papers see
M.~Rausch de Traubenberg,
arXiv:hep-th/9802141.

\bibitem{Filippov}
A.~T.~Filippov and A.~B.~Kurdikov,
hep-th 9312081

\bibitem{GL}
M.~Chaichian and A.~P.~Demichev,
{\it Introduction To Quantum Groups},
(Singapore, World Scientific, 1996).

\bibitem{Azcarra} J. A. Azcarraga, P.P. Kulish and F. Rodenas,
Czech. J. Phys. {\bf 44} (1994) 981 and Fortsch. Phys. {\bf 44} (1996) 
1-40.

\bibitem{Mepavi} M. M\'ezard, G. Parisi and M. A. Virasoro,
{\it Spin-glass theory and beyond}, (World Scientific,
Singapore, 1987).

\bibitem{Efetov} K. B. Efetov, Adv. Phys. {\bf 32} (1983) 53.
{\it Supersymmetry in disorder and chaos} (Cambridge
Univ. Press, Cambridge, 1997).

\bibitem{jorge} See, for instance, J. Kurchan
{\it Supersymmetric, replica and dynamic treatments
of disordered systems: a parallel presentation}, to appear
in Journal of Markov processes and related fields, cond-mat/0209399.


\end{thebibliography}
\end{document}